# Excluded volume effect in flexible dendrimer systems: A self-consistent field theory

Meng Shi[a], Yingzi Yang*[,a], and Feng Qiu[a]

[a] State Key Laboratory of Molecular Engineering of Polymers, Key Laboratory of Computational Physical Science, Department of Macromolecular Science, Fudan University, Shanghai 200433, China

We have studied the conformational and scaling behaviors of a flexible dendrimer immersed in athermal or good solvents. A self-consistent field theory combined with a pre-averaged excluded volume potential representing the two-body short-ranged interaction between the segments, was adopted to calculate the density profile of various generations and branch points thoroughly. Our calculation results support the "dense-core" model. We find the conformation of the dendrimer is strongly stretched in the dense central region, but much weakly stretched in the outer region where the segment density profile is shoulder shaped. Both our self-consistent field theory calculation and the Flory mean-field theory calculation give the same scaling law $R \sim (GP)^{1/5} N^{2/5}$, where $G$ is the generation number of the dendrimer, $P$ is the spacer segment number, and $N$ is the total segment number. If we fix $G$, the scaling law is simplified to $R \sim P^{3/5}$ in good solvent.

**Keywords**   dendrimer, self-consistent field theory, scaling law

## Introduction

Dendrimers are highly branched polymers synthesized generation by generation regularly. Since first prepared in 1978 by Vögtle,[1] dendrimer has been extensively studied both experimentally[2-9] and theoretically[10-27] over the past decades for its potential applications in various areas, including sensing, catalysis, molecular electronics[28], biomedicine,[28,29] and drug delivery system,[30,31] etc. In order to improve the performance of dendrimer in different areas, the physical property is crucial. Especially, the physical property of dendrimers in solutions attracts quite a lot of attentions, because for many applications the dendrimers are immersed in a solvent. However, due to the complex molecular structure and excluded volume effect, the entire conformational property and scaling laws have not yet reached a complete consensus.

(1) **Density profile**: "hollow-core" model and "dense-core" model

An early study on the structure of dendrimers was performed by de Gennes and Hervet in 1983.[10] They adopted a self-consistent field theory based on the assumption that "near the center the spacers may behave like flexible coils, but in the outer region they must be elongated".[10] Thus, the density profile of the dendrimer is minimum at the center and increases towards the edge. This is known as the "hollow-core" model (or "dense-packing" model). This model dominated the understanding on dendrimer for two decades and stimulated a lot application ideas,[28] because the model implies a molecular capsule to packing functional drugs for delivery.

However, in 1990, in contrast to the "hollow-core" model, Lescanec and Muthukumar reported a "dense-core" model based on a computer simulation,[11] in which the density profile of the dendrimer is maximum at the center. After that, a great deal of works based on numerical simulation and calculation methods, such as Monte Carlo (MC) simulation,[12-18] molecular dynamics (MD),[2,19,20] Brownian dynamics (BD),[21,22] self-consistent field theory (SCFT),[23,24] and density functional theory (DFT),[32] were published supporting the "dense-core" model. The results of different approaches agree on several features of single dendrimer system. First, the spacers with a small generation number are quite localized and contribute to a high density peak at the center of dendrimer. Second, with the increase of the generation number, a density plateau emerges besides the center density peak. This indicates that the dendrimer with small generation number is at a dilute state in good solvent, while the dendrimer with high generation number possess a dense state. The experiments of small-angle X-ray scatterings (SAXS) and small-angle neutron scattering (SANS) confirm the "dense-core" model.[4, 6,33]

Although the "dense-core" model is widely accepted nowadays, a local minimum valley between the center peak and the plateau region in the density profile, first discussed by Mansfield and Klushin,[12] remains confusing. This minimum is found in all simulations with strong steric interactions when the generation number $G$ is large enough.[2,13,16,17,19] However, the minimum is not found in Boris and Rubinstein's SCFT calculations,[23] where the segment distribution of the dendrimer is assumed to be proportional to a single branch chain in the presence of a mean density filed due to all of the monomers in the molecule.

---

*   E-mail: yang_yingzi@fudan.edu.cn

(2) **Conformation**: the back-folding phenomenon and the terminal segment distribution

According to the "hollow-core" model, surface congestion occurs at the periphery of a dendrimer after a certain generation. The spacers are strongly stretched outwards, and consequently the free ends of the last generation distribute near the shell. However, the simulation works supporting "dense-core" model found the terminal segment spread through the whole dendrimer region.[12-14,17-19,34] This indicates the strong back-folding tendency of the arms of the dendrimer. This phenomenon is confirmed by SANS experiments.[6]

(3) **Scaling laws (with solvent)**

The controversy over the scaling laws between the radius of gyration $R_g$ and other topological parameters of dendrimers, such as the segment number $N$, the spacer length $P$ and the generation number $G$, never stops.

The theoretical approaches predict different power laws in different models. In the "hollow-core" model, De Gennes and Hervet reported $R_g \sim N^{0.2}$ in good solvent (athermal).[10] In the "dense-core" model, via a mean-field, Flory-type approach based on the free energy of the linear branch,[15,23,34] the scaling law is $R_g \sim N^{1/5}(PG)^{2/5}$ in good solvent and $R_g \sim N^{1/4}(PG)^{1/4}$ in $\Theta$ solvent. In the poor solvent limit, the dendrimer is a dense sphere, so that $R_g \sim N^{1/3}$.

**Table 1** Summary of scaling laws of dendrimer in solvents.

| Ref | Method | Scaling law | Solvent |
|---|---|---|---|
| [7] | SAXS | $R_g \sim N^{0.42}$ with small $G$ <br> $R_g \sim N^{0.21}$ with large $G$ | polyamidoamine in methanol |
| [8] | SAXS | $R_g \sim N^{1/3}$ | polyamidoamine in methanol |
| [11] | MC | $R_g \sim (GP)^{0.57 \pm 0.05}$ <br> $R_g / P^{0.5} \sim N^{0.22 \pm 0.02}$ with high $N$ | good |
| [15] | MC | $R_g \sim (GP)^{2/5} N^{1/5}$ <br> $R_g \sim N^{0.3}$ | good <br> poor |
| [16] | MC | $R_g \sim N^{1/3}$ -small $G$ <br> $R_g \sim N^{0.24}$ -large $G$ | good <br> good |
| [19] | MD | $R_g \sim N^{0.3}$ with $P = 7$ | good, $\Theta$, poor |

In computer simulations and experiments, the scaling laws also do not agree with each other, as briefly summarized in Table 1. A more detailed table summarizing different power law exponents in theoretical works and simulations can be found in ref. [18].

In order to solve the problems listed above, we perform the SCFT based on the full molecule instead of the linear branch chain,[23] and study the conformations and scaling laws of a flexible dendrimer in good solvent thoroughly. The SCFT based on a linear branch chain for dendrimer system, first introduced by Boris and Rubinstein,[23] adopts a pre-averaged potential to represent the two-body interaction between the segments. In Flory theory,[35] this potential is interpreted as the effect of volume exclusion between the segments of linear chains in a good solvent. Comparing with MD and MC with full steric interaction, SCFT with the pre-averaged, two-body interaction potential can give good expression of the segment density profile and the power laws in good solvents, although it is hard to approach the high density limit where three-body interaction plays an important role. We study the dendrimers with the functionality of the center segment and the branch segments equals to three, which is called the standard dendrimers.[36]

Our paper is organized as follows. In Sec. II we introduce the dendrimer model and the self-consistent field theory with excluded volume potential, and discuss the Flory theory for the dendrimer. In Sec. III, we focus on the physical aspects (density profile, folding-back conformation, stretching conformation and power laws) of a dendrimer immersed in a good solvent via a detailed analyzing of our simulation data. In the last section, we summarize the main results and draw the conclusion.

## Theoretical method

### A. The model

We consider a dendrimer of $G$ generations with the functionality of the center segment $f_0$ and the functionality of the branch segments $f$, as illustrated in Fig.1. The first segment is fixed at the origin. We use $g$ to denote the generation number of the spacers. The branch segment, denoted with a number $p$, connects the $g$-th and the $(g+1)$-th spacer. The first segment and the terminal segments are denoted as $p=0$ and $p=G$, respectively. For simplicity, all the spacers, which are assumed to be flexible chains, have the same segment number $P$. The total segment number is

$$N = \begin{cases} 1 + f_0 GP, & f = 2 \\ 1 + f_0 P \dfrac{(f-1)^G - 1}{f-2}, & f > 2 \end{cases} \quad (1)$$

In the absence of excluded volume interactions, or in other words, when the dendrimer is immersed in a $\Theta$ solvent, the conformation entropy achieves the maximum value with each linear side chain (a no-backward route from $p=0$ to one end segment $p=G$) assuming Gaussian conformation. Therefore, $R_0$, the square root of the mean square distance from the center segment $p=0$ to a terminal segment $p=G$, has the same form as the mean end-to-end distance of a linear Gaussian chain with segment number $PG$ [15,23]

$$R_0 = (PG)^{1/2} a \quad (2)$$

where $a$ is the statistical segment length. In Sec. III A, we confirm this power law by SCFT calculation.

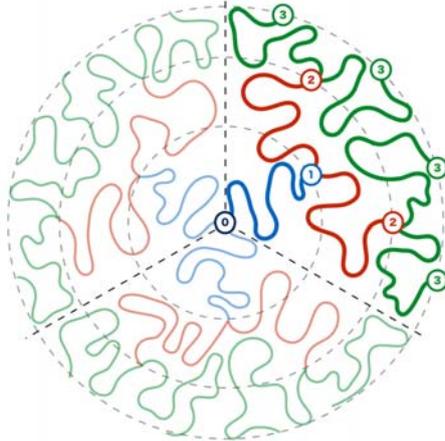

**Figure 1** Illustration of the structure of a standard dendrimer with $G=3$, $f_0 = f = 3$. The blue, red, green lines represent the spacers with generation number $g=1$, $g=2$, and $g=3$, respectively. The circles denote the branch segments and the free end segments, and the numbers inside the circle are corresponding $p$ values. The sharp colored area represents a dendron, which is one-third of a dendrimer.

### B. Flory mean-field theory

We consider the dendrimer immersed in a good solvent, and assume that the segment density in the spatial area occupied by the dendrimer is not dense. The conformation of the polymeric molecule is determined by the balance between the repulsive excluded volume interaction and the conformational entropy. Based on the idea of Flory free energy for a single linear chain in good solvent,[35] we write the free energy of a dendrimer immersed in a good solvent as,

$$\frac{F}{k_B T} \sim \frac{R_g^2}{R_{g0}^2} + u \frac{N^2}{R_g^3} \quad (3)$$

where $k_B$ is the Boltzmann constant, $T$ is temperature, $u$ is the mean-field approximated, excluded volume parameter, "$\sim$" means "is proportional to", and the prefactors are neglected here. The radius of gyration of a phantom Gaussian dendrimer $R_{g0}$ is,[36]

$$R_{g0} \sim (GP)^{1/2} a \quad (4)$$

when $G \gg 1$.

Comparing Eqn. 3 and the pioneer mean field (MF) calculations,[16,18,34] we modified the calculation in several as-pects listed as below.

(1) The free energy Eqn. 3 is based on the whole dendrimer molecule instead of a linear side chain. Therefore, we assume that the entropic energy of dendrimer changes proportional to $R_g^2 / R_{g0}^2$. For the second term on the right side of Eqn. 3, we replace the linear side chain length $PG$ in the pioneer MF calculations with the total segment number $N$.[15,23,34]

(2) Neglect the three-body interaction term in Sheng et al.[15] We assume the segment density low enough that the three-body interaction is insignificant.

Minimizing the free energy, we obtain a scaling law of the equilibrium size in terms of $N$, $PG$ and $u$,

$$R_g \sim u^{1/5}(PG)^{1/5} N^{2/5} \qquad (5)$$

where $u$ is a phenomenological constant to describe the solvent property. We drop $u$ and simplify the power law as $R_g \sim (PG)^{1/5} N^{2/5}$.

## C. Self-consistent field theory

SCFT was first introduced by Edwards and developed by Matsen[37] and Fredrikson et al.[38] It has been used successfully to study the thermodynamic equilibrium of polymers in its present form. However, the SCFT theory usually does not include the excluded volume interaction, which is absolutely important in dendrimer systems of high generation. Therefore, we write the pair interaction term $U$ to represent the averaged volume exclusion interaction between the segments,[39]

$$U = \frac{u}{2v_0}\int d\mathbf{r}\left[\hat{\Phi}(\mathbf{r})\right]^2 = \frac{1}{v_0}\int d\mathbf{r}\omega(\mathbf{r})\hat{\Phi}(\mathbf{r}) \qquad (6)$$

where $v_0 = 1$ is the segment size, and $\omega(\mathbf{r})$ is the pre-averaged external field exerted on the dendrimer to represent the volume exclusion effect. The segment number density operator $\hat{\phi}(\mathbf{r})$ is defined as,

$$\hat{\Phi}(\mathbf{r}) = v_0 P \sum_\alpha \int_0^1 ds\, \delta(\mathbf{r} - \mathbf{r}_\alpha(s)) \qquad (7)$$

where $\mathbf{r}_\alpha(s)$ is the position vector of the $s$-th segment on the $\alpha$-th spacer. The parameter $u$ describes the overall effect of the volume exclusion of the segments and the swelling of the solvent. At $\Theta$ point with $u = 0$, where a slight incompatibility of solvent balances the excluded volume effect of segments, the property of dendrimer is similar to the ideal Gaussian case. As $u$ increases, the size of the dendrimer expands due to the excluded volume interaction.

For a given conformation, the Hamiltonian of the dendrimer system is given by

$$\frac{H}{k_B T} = \frac{3}{2Pa^2}\sum_\alpha \int_0^1 ds \left|\frac{d\mathbf{r}(s)}{ds}\right|^2 + \frac{1}{v_0}\int d\mathbf{r}\omega(\mathbf{r})\hat{\Phi}(\mathbf{r}) \qquad (8)$$

So the partition function of the dendrimer is written as

$$Z = \int \prod_\alpha D\mathbf{r}_\alpha(s)\exp\left[-\frac{H(\mathbf{r}(s))}{k_B T}\right]\cdot \Delta \qquad (9)$$

where $\Delta$ function represents the connection restrictions at the branch segments and the center segment. By inserting $\int D\Phi \delta(\Phi(\mathbf{r}) - \hat{\Phi}(\mathbf{r})) = 1$ and transfer Eqn. 9 from the particle description to a field description, the partition function is rewritten as

$$Z = \int D\Phi(\mathbf{r})\int DW(\mathbf{r})\exp\left\{-\frac{F[\Phi,W]}{k_B T}\right\} \qquad (10)$$

and the free energy functional

$$\frac{F[\Phi,W]}{k_B T} = -\ln Q - \frac{1}{v_0 P}\int d\mathbf{r}[W(\mathbf{r}) - P\omega(\mathbf{r})]\Phi(\mathbf{r}) \qquad (11)$$

$$Q = \int \prod_\alpha D\mathbf{r}_\alpha \exp\left\{-\int_0^1 ds W(\mathbf{r}_\alpha(s)) - \frac{3}{2Pa^2}\int_0^1 ds\left|\frac{d\mathbf{r}_\alpha(s)}{ds}\right|^2\right\}\cdot \Delta \qquad (12)$$

The quantity $Q$ is the partition function of the dendrimer in an (imaginary) external field $W(\mathbf{r})$. Because the functional integral $\int D\Phi \int DW$ is hard to calculate exactly, we therefore adopt the saddle point approximation which takes the most possible configuration as the whole integral. Then, the self-consistent field equations describing the equilibrium behavior of the system are

$$\Phi(\mathbf{r}) = -\frac{v_0 P}{Q} \frac{\partial Q}{\partial W} = \frac{v_0 P}{Q} \sum_\alpha \int_0^1 ds\, q_\alpha(\mathbf{r},s) q_\alpha^+(\mathbf{r},s) \quad (13)$$

$$W(\mathbf{r}) = \frac{uP}{2} \Phi(\mathbf{r}) \quad (14)$$

where $q_\alpha(\mathbf{r},s)$ is the so-called propagator which represents the probability of finding the $s$-th segment on the $\alpha$-th spacer at position $\mathbf{r}$ in space. $q_\alpha(\mathbf{r},s)$ and $q_\alpha^+(\mathbf{r},s)$ satisfy the modified diffusion equations

$$\frac{\partial}{\partial s} q_\alpha(\mathbf{r},s) = \left[\frac{Pa^2}{6} \nabla^2 - W(\mathbf{r})\right] q_\alpha(\mathbf{r},s) \quad (15)$$

$$\frac{\partial}{\partial s} q_\alpha^+(\mathbf{r},s) = -\left[\frac{Pa^2}{6} \nabla^2 - W(\mathbf{r})\right] q_\alpha^+(\mathbf{r},s) \quad (16)$$

with the initial condition defined by the connection restriction function $\Delta$, and the center segment is fixed at the origin point of the space.

Note that, when two spacers have the same generation number $g$, the solvent of their diffusion equations are the same. Therefore the segment density of the dendrimer can be written as,

$$\Phi(r) = \frac{v_0 P}{Q} f^0 \sum_{g=1}^G (f-1)^{g-1} \int_0^1 ds\, q_g(\mathbf{r},s) q_g^+(\mathbf{r},s) \quad (17)$$

where $q_g(\mathbf{r},s)$ represents the propagator of the spacers with generation number $g$.

The numerical calculation process is performed by initiating an external field $W(\mathbf{r}) = 0$, solving the diffusion equations (Eqn. 15 and Eqn. 16), calculating the dimensionless density Eqn. 17, generating the new field according to Eqn. 14, then iterating from the diffusion equation again. Taking the advantage of spherical symmetry of this model, we could reduce the diffusion equation from three dimension into one dimension along the radius $r$. In spherical coordinate, Eqn. 15 could be rewritten as

$$\frac{\partial}{\partial s} q_g(r,s) = \frac{Pa^2}{6} \frac{1}{r^2} \frac{\partial}{\partial r} r^2 \frac{\partial}{\partial r} q_g(r,s) - W(r) q_g(r,s)$$

$$(18)$$

In the present work, we adopt the implicit method to solve Eqn. 15 with the boundary condition

$$\left.\frac{\partial q_g(r,s)}{\partial r}\right|_{r=r_{max}} = 0 \quad (19)$$

for any $g$ and $s$, where $r_{max}$ is the radius of the calculation space.

Finally, we define the mean-squared radius of the dendrimer as,

$$\langle R^2 \rangle = \frac{1}{N} \int dr\, 4\pi r^4 \Phi(r) \quad (20)$$

and the mean-squared center-to-branch-segment distance,

$$\langle R_p^2 \rangle = \frac{1}{f_0(f-1)^{p-1}} \int dr\, 4\pi r^4 q_p(r,1) \quad (21)$$

When $p = G$, $R_p$ is the center-to-terminal distance.

In this paper, we choose the statistical segment length $a$ and the segment volume $v_0$ as units. For simplicity, the functionality of the branch segments are fixed to be $f = 3$. The functionality of the center segment is $f_0 = 3$ for a standard dendrimer, and is $f_0 = 1$ for a dendron.

## Results and discussion

In this section, we exhibit the SCFT calculation results of dendrimer systems by systematically changing the parameters $G$, $P$, and $u$.

We first analyze the end-to-end distance $R^2$ of a linear chain with different $u$, in order to find a proper value of $u$ to ensure the correct scaling behavior $R^2 \sim P^\beta$.

Then, we analyze the density profiles of the dendrimer with the proper $u$ value to show the swelling behavior of a dendrimer in a good solvent. We focus on the density profiles of a single dendrimer, because the segment density not only reflects the size and the conformation of the molecule, but also closely relates to the dynamic properties of the solution such as viscosity. Moreover, one can design the function of a dendrimer if the accurate knowledge of the spatial position of each segment is known.

At last, we compare the scaling laws obtained from SCFT and Flory mean-field theory. The scaling law, $R \sim (GP)^{1/5} N^{2/5}$, predicted by Flory mean-field theory in our work is based on the full molecule. Therefore, it is different from the scaling law $R^2 \sim (GP)^{2/5} N^{1/5}$ predicted by the Flory mean-field theory based on the linear side chain.[15,23,34] We confirm via our SCFT that our scaling law $R \sim (GP)^{1/5} N^{2/5}$ is a proper description for the dendrimer with flexible spacers in the good solvent, where the three-body interaction can be neglected.

### A. Linear Chain with $G = 1$ and $f_0 = 1$

With $G = 1$ and $f_0 = 1$, our model represents a linear chain with one end anchored on the origin of the space. The mean squared end-to-end distance $\langle R_1^2 \rangle$ of the linear chain has the power law relation to the segment number $P$ when $P \gg 1$,

$$\langle R_1^2 \rangle \sim P^\beta \qquad (22)$$

The exponent of the scaling law, $\beta$, depends on the interaction between the segments in our model.

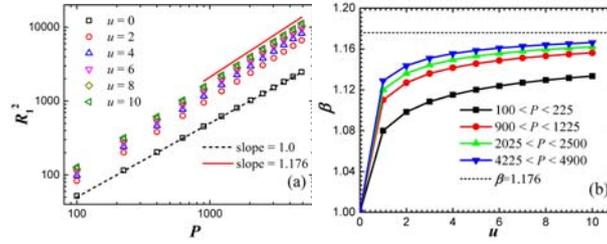

**Figure 2** (a) The Scaling law relation $R_1^2 \sim P^\beta$ of a single chain with different $u$. (b)The exponent of the scaling law relation in (a) as a function of $u$ within different $P$ value regions. The dashed line is $\beta = 1.176$.

We plot $\langle R_1^2 \rangle$ as a function of $P$ in Figure 2(a). When $u = 0$, our model represents a phantom Gaussian chain, so that the data exhibit a perfect power law relation $\langle R_1^2 \rangle = Pa^2$. The Gaussian chain is the simplest chain model that the "ghost" segments do not interact via volume exclusion. When $u \neq 0$, the repelling force from between the segments extends the chain coil, while the conformational entropy tends to pull the coil size back to the Gaussian coil size. The balanced $\langle R_1^2 \rangle$ increases with $u$, and the power law exponent $\beta$ increases, as shown in Figure 2(b). $\beta$ starts at value 1 with $u = 0$ for the Gaussian chains, and increase rapidly with $u < 2$. The further increasing of $u$ only causes a smooth increase of $\beta$, and apparently $\beta$ approaches a limit value $\beta \approx 1.176$ with increasing $u$ and $P$. Therefore, with the mean-field-approximated volume exclusion parameter $u$, the SCFT can reproduce the scaling behavior of the mean-squared end-to-end distance of the self-avoid-walking chain when $u$ and $P$ are large enough.

Due to the limit of the computational labor, we use $u = 4$ and $P = 1000$ in our calculation of the dendrimer systems. As shown in Figure 2(b), $\beta \approx 1.14$ with $u = 4$ and $P = 1000$, which ensures the error from the limit value $\beta = 1.176$ less than 5%.

### B. Density profiles of dendrimer

The total segment number density of the dendrimer, $\Phi(r)$, is shown in Fig. 3(a). Because the center segment is fixed at the origin point, $\Phi(r)$ has a high peek at $r = 0$. After the sharp decreasing from the peek, $\Phi(r)$ exhibit a "shoulder" region with very slow decreasing rate. The width of the shoulder increases rapidly with increasing $G$. However, although the total segment number $N$ increases exponentially with $G$, the value of $\Phi(r)$ in the shoulder region decreases. Beyond the shoulder region, $\Phi(r)$ drops rapidly. The monotonically decreasing density profile in the "shoulder" region supports the "dense-core" model for dendrimers in good solvent.[11,13-14,16,17,19]

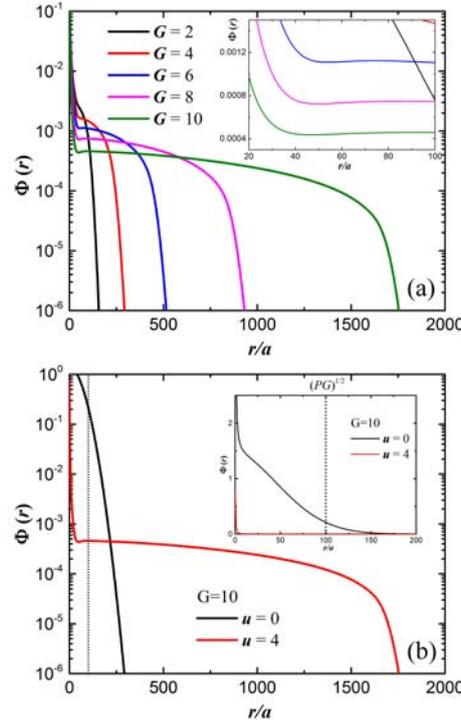

**Figure 3** (a) The total segment number density $\Phi(r)$ of the dendrimers with $u=4$ and $P=1000$. The insert shows the shallow valley region with $G \geq 6$. (b) $\Phi(r)$ with $G=10$ and $u=0, 4$. The dashed line indicates the root of the mean-squared end-to-end distance of a Gaussian chain with $PG=10^4$

Note that, there is a shallow valley between the peek and the "shoulder" region with $G \geq 6$, as shown in the insert in Figure 3(a). The local minimum was also observed in MC and MD simulations with full volume exclusions, when the solvent is good, the spacer is short, or $G$ is large enough.[17-19]

Compared with the dendrimer with $u=0$ (the Gaussian dendrimer), the dendrimer with $G=10$ and $u=4$ exhibits the obvious swelling behavior, as shown in Figure 3(b). The radius of the Gaussian dendrimer is determined by $(GP)^{1/2}$. When $u>0$, the volume exclusion interaction enlarges the dendrimer size, while the conformational entropy of the dendrimer prefers the Gaussian conformation. The shoulder region of $\Phi(r)$ emerges due to the balance between the volume exclusion and the conformational entropy. The emergence of the shoulder agrees the result of MD and MC simulations with full volume exclusion.[13,14,16,17,19]

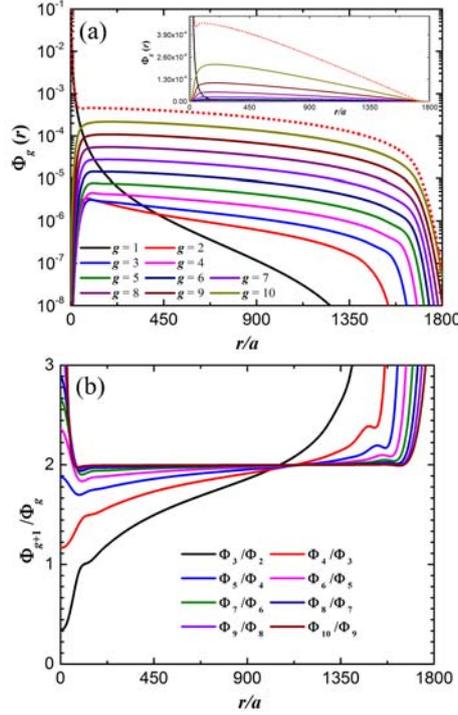

**Figure 4** (a) The segment number density $\Phi_g(r)$ with $g = 1 \sim 10$ of the dendrimer with $G = 10$, $P = 1000$ and $u = 0.01$. The red dotted line is the total segment density $\Phi(r)$. The insert is the same diagram with linear axis. (b) The fraction of the segment number density of adjacent generations, $\Phi_{g+1}(r)/\Phi_g(r)$.

In order to clarify the contribution of different parts of the dendrimer to the total segment distribution, we analyze the segment number density $\Phi_g(r)$ of the $g$-th generation spacers. We take the dendrimer with $G = 10$ as an example, as shown in Fig. 4(a). The segment density of the first generation, $\Phi_1(r)$, mainly contributes to the center peak of the total segment number density $\Phi(r)$, because the center segment is fixed at $r = 0$. $\Phi_g(r)$ with $g \geq 2$ appears to be uniform in the shape of the curves. The density profiles $\Phi_{g\geq 2}(r)$ approach zero at $r = 0$, because the center region is crowded of the first-generation segments. In the shoulder region, $\Phi_{g\geq 2}(r)$ decrease smoothly with $r$ until a limit distance where they decrease rapidly. The influence of the fixed center segment decays with the generation number $g$. As shown in Figure 4(a), $\Phi_2(r)$ has the elevated peak close to the center, while $\Phi_{g\geq 3}(r)$ appears to be relatively "parallel" to each other in the shoulder region.

In Fig. 4(a), the segments with small generation numbers are fairly localized ($g = 1$ and 2 in our model), while the density of the segments at higher generations ($g > 2$ in Figure 4(a)) at the shoulder region increases with $g$. We plot $\Phi_{g+1}/\Phi_g$ in Figure 4(b), and find that $\Phi_{g+1}/\Phi_g$ is always greater than 1 except for $g = 2$. Therefore, the probability to find a segment on the spacer with larger $g$ is always higher than that with smaller $g$ for $g \geq 3$. Moreover, in the shoulder region, $\Phi_{g+1}/\Phi_g \approx 2 = f - 1$ with $g \geq 5$. Therefore, for a dendrimer with large $G$, $\Phi_g$ is approximately proportional to the total segment number of the generation, i.e., $\Phi_g \propto (f-1)^{g-1}$ with $g \geq 5$.

This is similar to the result obtained by Murat and Grest.[19] In their MD simulations in athermal solvent, the generations $1 \leq g \leq 5$ are localized, with the peak of their density profiles moving farther away from the core with increasing $g$. And at the higher generations, the density of the segments in the shoulder region increases with $g$. However, in our model, the spacers are extremely flexible, thus only the spacers with $g = 1$ and $g = 2$ are obviously localized.

Note that, in Figure 4(a), $\Phi_{10}(r)$ has higher value than other $\Phi_g(r)$ near the center peak region except for $\Phi_1(r)$. This indicates the back-folding conformation. The spacers with large $g$ can spread to larger distance, and at the same time they can fold back due to the conformational entropy. Although the volume exclusion interaction swells the dendrimer extensively, it is inadequate to avoid the back-folding conformation. The back-folding phenomenon was reported in previous experimental[6] and theoretical works.[2,15,16,18,34]

## C. The stretched conformation

The equilibrium conformation of a dendrimer is determined by the compromise between the conformational entropy and the volume exclusion interaction. The flexible spacers stretch to decrease the local segment density and consequently suppress the two-body interaction energy, in the cost of increasing the entropic energy due to the deviation from the ideal Gaussian dendrimer.

The stretching of a chain can be evaluated by its end-to-end distance. Therefore, we calculate the number density of the branch points, $\phi_p(r)$, as shown in Figure 5. Then we obtain the mean-squared center-to-branch-segment distance $\langle R_p^2 \rangle$, and compare it with the Gaussian mean-squared end-to-end distance $pPa^2$, as shown in Figure 6.

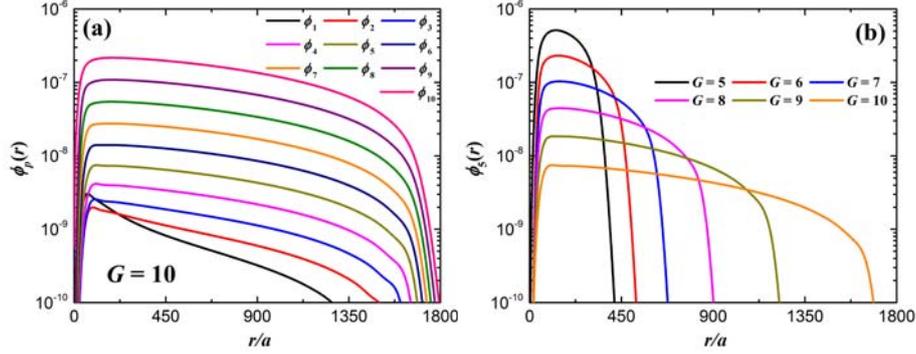

**Figure 5** (a) The number density of the $p$-th branch segment, $\phi_p(r)$ of the dendrimer with $G=10$, $P=1000$, and $u=4$. (b) $\phi_5(r)$ of the dendrimers with $P=1000$, and $u=4$.

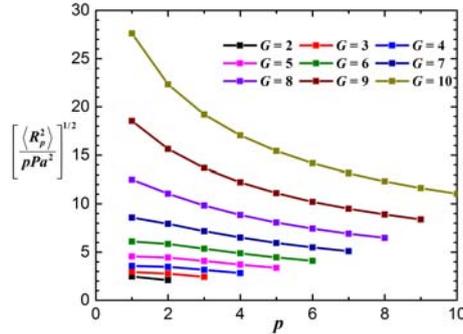

**Figure 6** The mean-squared center-to-branch-segment distance $\langle R_p^2 \rangle$ normalized by the Gaussian mean-squared end-to-end distance $pPa^2$. The calculation parameters are $P=1000$ and $u=4$.

$\phi_0(r)$ is the Dirac delta function at $r=0$, which is not exhibited in the figure. As shown in Figure 5(a), $\phi_p(r)$ expends wider with larger $p$. Same to the phenomenon for $\Phi_g(r)$ in Figure 4(a), the branch points with $p=1$ is quite localized. $\phi_1(r)$ exhibits a lifted peak near the center region. On the other hand, for the dendrimers with different $G$, $\phi_p(r)$ expends wider with lower value in the shoulder region with larger $G$, as shown in Figure 5(b).

The mean squared center-to-branch-segment distance $\langle R_p^2 \rangle$ normalized by $pPa^2$ is used to evaluate the average degree of stretching of the side chain connecting the center segment and the $p$-th branch segment, as shown in Figure 6. The spacers are more stretched with larger $G$, because the total segment number increases exponentially with $G$, and the entropic spacer springs extend farther to balance the stronger volume exclusion interaction. $\langle R_p^2 \rangle / pPa^2$ decreases with $p$, which indicates that the stretching behavior is inhomogeneous along the linear side chain inside a dendrimer. The spacers with smaller $p$ stretches more strongly than the spacers with larger $p$. Therefore, near the center region, the spacers are elongated, while in the outer region, the spacers are more relaxed.

Note that $\langle R_p^2 \rangle$ is analyzed in the radial direction in the sphere coordination. Therefore, we only discuss the radial stretching of the dendrimer conformation, while the deformation of the conformation in longitude and latitude directions are not of focus here.

**D. Scaling law analysis**

Via the Flory mean-field theory approach based on the full dendrimer, we obtain the scaling relation $R_g \sim (PG)^{1/5} N^{2/5}$, as shown in Eqn. 5. Since $N$ is proportional to $P$, we obtain $R_g \sim P^{3/5}$ when $G$ is fixed. The scaling law $R_g \sim P^{3/5}$ agrees with pioneer researches on the scaling law for good solvent.[11,15,18] However, via a mean-field, Flory-type approach based on the free energy of the linear branch,[15, 34] the scaling law is, $R_g \sim (PG)^{2/5} N^{1/5}$, which is different from our MF estimation in Equation 5 by exchanging the exponents on $N$ and $(PG)$. In this section, in order to validate the scaling laws, we evaluate the size of the dendrimer by calculating the mean-squared radius the segments from the center, $\langle R^2 \rangle$. Since the center segment is fixed at the origin, $\langle R^2 \rangle$ is not exactly equal to $\langle R_g^2 \rangle$ of a dendrimer in dilute condition. Nevertheless, we consider that the confinement of one segment does not strongly influence the size of the molecule. Thus, $\langle R^2 \rangle$ has the same scaling relation laws as $\langle R_g^2 \rangle$ does.

The mean squared radius $\langle R^2 \rangle$ of the dendrimers with fixed $G$ is plotted as a function of the spacer length $P$ in Figure 7. For the dendrimers with $G = 2 \sim 8$, we obtain $\langle R^2 \rangle \sim P^\beta$, where $\beta$ increases from 1.17 with $G = 2$ to 1.20 with $G = 8$. $\beta$ appears to be convergent to a limit value. Since we choose $u = 4$ with the acceptable 5% deviation from the scaling law of the linear chain, we estimate the same error bar for $\beta$. Thus, we obtain $\langle R^2 \rangle \sim P^\beta$ with $\beta = 1.20 \pm 0.06$, which agrees with our MF theory prediction $\langle R^2 \rangle \sim P^{6/5}$ and other simulations.[11,15,18]

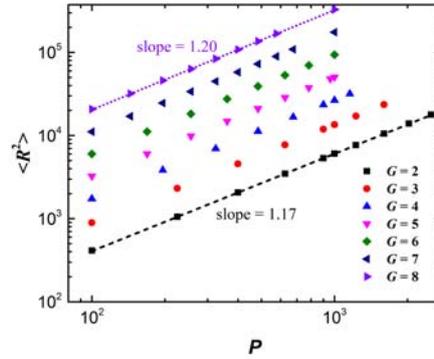

**Figure 7** The scaling relation $\langle R^2 \rangle \sim P^{6/5}$ of dendrimers with $G = 2 \sim 8$ and $u = 4$.

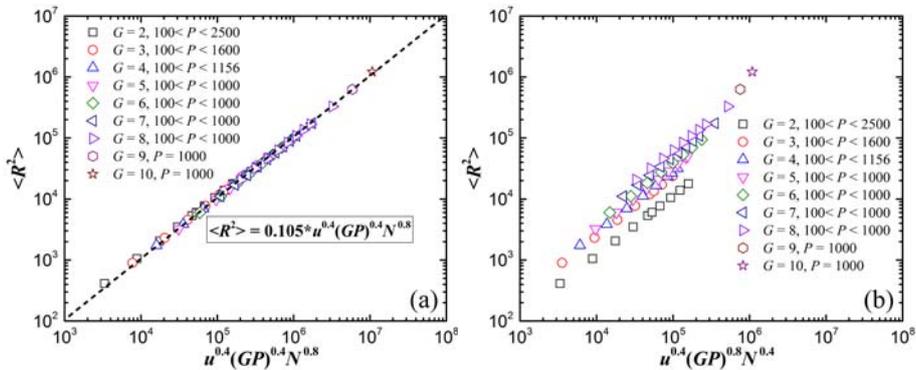

**Figure 8** The scaling relation (a) $\langle R^2 \rangle \sim (GP)^{2/5} N^{4/5}$, and (b) $\langle R^2 \rangle \sim (GP)^{4/5} N^{2/5}$ of the dendrimers with $G = 2 \sim 10$, $u = 4$, and varied $P$. The insert equation in (a) is the linear fitting of all data.

Then, we plot $\langle R^2 \rangle$ as a function of $u^{2/5}(PG)^{2/5} N^{4/5}$ and $u^{2/5}(PG)^{4/5} N^{2/5}$, as shown in Figure 8. With

$2 \leq G \leq 10$ and $100 \leq P \leq 2500$, the range of $\langle R^2 \rangle$ covers at least four magnitudes. In Figure 8(a), the linear fitting function $\langle R^2 \rangle = 0.105 u^{2/5} (GP)^{2/5} N^{4/5}$ fits all the data well. Our full-dendrimer-based scaling law obtained via the Flory mean-field theory approach agrees well with the SCFT results. However, when the horizontal axis is $u^{2/5} (PG)^{4/5} N^{2/5}$, the data points are not focused on the same line. Although the data points with same $G$ seems to have linear relation in Figure 8(b), the slopes of the fitting lines for differnt $G$ are diverse. Therefore, we verify that the full-dendrimer-based scaling law $R_g \sim (GP)^{1/5} N^{2/5}$ is validate for the dendrimers with long and flexible spacers in a good or athermal solvent.

In the MC simulations,[15] the data agrees quite well with the side-chain-based scaling law $R_g \sim (GP)^{2/5} N^{1/5}$ with $N < 2000$. We notice that the volume exclusion in MC simulations is the steric interaction, while in SCFT we employ the mean field approximation to represent the two-body volume exclusion interaction. Moreover, the spacer length in MC simulations ($1 \leq P \leq 50$) is much smaller than that in our SCFT calculation ($100 \leq P \leq 2500$). Therefore, the disagreement on the scaling behavior between MC simulations and our SCFT calculation is raised from the details of the model. The model in MC simulations describes the small dendrimers with short spacers. The model in SCFT describes the large dendrimers with long, flexible spacer chains, where the segment density inside the dendrimer is very low, and the volume exclusion interaction is dominated by two-body interaction.

## Summary and Conclusions

In summary, we investigate the conformational and scaling behaviors of dendrimers with volume exclusion interaction in a good solvent by using Flory mean-field theory and SCFT method. Our SCFT approach takes into account the excluded volume effect by using a pre-averaged potential describing the two-body volume exclusion interaction between segments. The solvent is implicitly described in our SCFT.

Our SCFT results confirm the "dense-core" model of the dendrimers, accompanied by a persistent folding back conformation. In the smooth shoulder region of the segment density profile of a dendrimer with $G$ generations, the density of the segments on the $g$-th generation, $\Phi_g(r)$, is approximately proportional to their total number $f_0 (f-1)^{g-1}$. Even near the center region, $\Phi_{g+1}(r)/\Phi_g(r) > 1$, which indicates that the outer spacers can fold back to the center region. The strong volume exclusion interaction swells the molecule extensively, but it is not adequate to avoid the back-folding conformation of the flexible chains. By analyzing the center-to-branch-segment distance of the side chains, we find that different parts of a linear side chain of a dendrimer are under different stretching conditions. The center spacers with the generation number $g = 1$ are mostly elongated.

The scaling laws between the size of the dendrimers and other topological parameters, such as the segment number $N$, the spacer length $P$ and the generation number $G$, is an unfinished debate. Our SCFT result supports the scaling law predicted by the Flory mean-field theory approach based on the full dendrimer. With the good solvent assumption, and neglecting the three-body interaction, we find the dendrimer size $R$ has the scaling law relation $R_g \sim P^{3/5}$ and $R_g \sim (GP)^{1/5} N^{2/5}$, predicted by both SCFT and Flory mean-field theory.

## Acknowledgement

We are grateful for the financial support from the National Natural Science Foundation of China (21320102005, 21774026) and Ministry of Science and Technology of the People's Republic of China (2016YFA0203301).

# Entry for the Table of Contents

Page No.
**Excluded volume effect in flexible dendrimer systems: A self-consistent field theory**

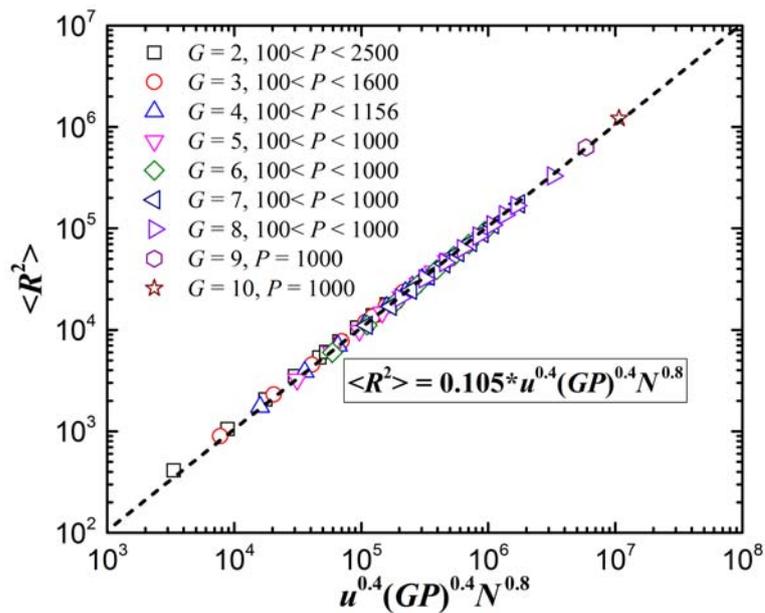

Meng Shi; Yingzi Yang*; Feng Qiu

The scaling law of the dendrimer with long, flexible spacers in a good solvent.